\documentclass{article}

 \usepackage{hyperref}
\usepackage{bm}
\usepackage{amssymb}	% Extra maths symbols
\usepackage{authblk}
\usepackage{graphicx}
\usepackage{tabularx, booktabs, makecell, caption}
\usepackage{siunitx}
\usepackage{float}
\usepackage[
singlelinecheck=false % <-- important
]{caption}

\begin{document}
\title{\bf Shadows and photon rings of a spherically accreting Kehagias-Sfetsos black hole}
\author{{Mohaddese Heydari-Fard$^{1}$%
\thanks{Electronic address: \href{mailto:m\_heydarifard@sbu.ac.ir}{m\_heydarifard@sbu.ac.ir}} ,  Malihe Heydari-Fard$^{2}$\thanks{Electronic address: \href{mailto:heydarifard@qom.ac.ir}{heydarifard@qom.ac.ir}} and Nematollah Riazi$^{1}$\thanks{Electronic address: \href{mailto:n\_riazi@sbu.ac.ir}{n\_riazi@sbu.ac.ir}}}\\ {\small \emph{$^{1}$ Department of Physics, Shahid Beheshti University, Evin 19839, Tehran, Iran}}
\\{\small \emph{$^{2}$ Department of Physics, The University of Qom, 3716146611, Qom, Iran}}}

\maketitle

\begin{abstract}
By considering Kehagias-Sfetsos black hole in the framework of the Ho\v{r}ava-Lifshitz gravity, we study the optical appearance of such black holes surrounded by spherical accretion flow. For the static/infalling spherical accretion flow, we compute the observed specific intensity as a function of impact parameter. We also investigate the effect of the Ho\v{r}ava parameter and accreting matter on the luminosity of shadows and photon rings. It is found that an increase in the Ho\v{r}ava parameter decreases the shadow size, while the shadows and photon rings luminosities increase. Moreover, we constrain the Ho\v{r}ava parameter from the observational data reported by the Event Horizon Telescope for M87* and Sgr A*.
\vspace{5mm}
\end{abstract}

\section{Introduction}
\label{1-introduction}
The Event Horizon Telescope (EHT) collaboration released the first image of a black hole shadow \cite{A1}--\cite{A7}. The Image is formed by null geodesics in the strong gravity regime. The photons with low angular momentum fall into the black hole and form a dark area for a distant observer, while photons with large angular momentum coming from infinity will be deflected by the gravitational potential of the black hole. However, the photons with critical value of angular momentum revolve around the black hole an infinite time and surround the dark interior, which are called the photon ring and the black hole shadow, respectively. In a seminal work, Synge calculated the angular radius of the shadow of Schwarzschild black hole \cite{Synge}. Then, Bardeen studied the shadow of Kerr black hole and argued that the black hole angular momentum causes the deformation of its shadow \cite{Bardeen}.

By modelling M87* with the Kerr geometry in general relativity (GR), the observation was found to be in agreement with the predictions of GR. However, due to the EHT systematic uncertainties it is still possible to test the alternative theories of gravity by simulating the black hole image and observing deviations from the Kerr solution. To this end, one can explore the distortion in the black hole image which contains valuable information about the structure of space-time around a black hole solution. This motivated many authors over the recent years to study the black hole shadow in the context of modified theories of gravity \cite{shadow2}--\cite{shadow13}.

On the other hand, the astrophysical black holes are expected to be surrounded by sources of the luminous accretion material which makes it possible to investigate the observational appearance of the black hole from the accretion flow. Indeed, before the discovery of the black hole shadow by EHT, the possible observational characteristics of the black hole shadow by considering different accretion flows were studied. Luminet was the first to investigate the optical properties of the Schwarzschild black hole in 1979, and constructed the simulated shadow image of the Schwarzschild black hole surrounded by an emitting thin accretion disk \cite{Luminet}. The simulated image obtained by Luminet is remarkably similar to the black hole shadow image captured by EHT \cite{A1}. He found that the emergence of the shadow and ring depends on the position and profile of the accretion flow and the inner edge of disk can have a remarkable signature in the image. Thereafter, Falcke et al. by considering the radiation of a hot optically thin accretion flow around a supermassive black hole in the center of our galaxy, created a ray-tracing code to obtain the images of Sgr A*, and showed that the black hole shadow is equivalent to the gravitational lensing effect \cite{Falcke}. For a geometrically thick and optically thin accretion disk, the gravitational lensing and the shadow of the Schwarzschild black hole was studied by Cunha et al. \cite{Cunha}. Gralla et al. by considering the Schwarzschild black hole with both thin and thick accretion disks, investigated the trajectory of light rays and ring that surrounds the black hole shadow. It was found that the shadow size depends on the position of the accretion disk as well as the emission profile of the model \cite{Gralla}. However, when the Schwarzschild black hole is surrounded by spherically symmetric accretion flow, Narayan et al. showed that the location of the outer edge of the shadow is independent of the inner radius at which the accreting gas stops radiating \cite{Narayan}. Therefore, the size of the shadow depends on the space-time geometry and is not affected by the details of the accretion flow. Also, the optical appearance of black holes surrounded by various accretions in modified gravity theories, have been extensively studied in \cite{a1}--\cite{a21}.

Amongst the many modifications of GR that have been suggested is the Ho\v{r}ava-Lifshitz gravity which is motivated by the need to include the quantum effects in the low-mass limit. The theory is a renormalizable four-dimensional theory of gravity which reduces to the Einstein's gravity with non-vanishing cosmological constant in the IR limit but with improved UV behavior. A class of static and spherically symmetric black hole solutions of the theory with a cosmological constant was obtained in \cite{s1}. Among them, the Ads type solution has an asymptotic behavior which differs from the Schwarzschild-Ads solution in GR; namely in the IR limit the theory of GR is not always recovered. However, in the context of the modified Ho\v{r}ava model, a static spherically symmetric solution with asymptotically flat behavior, which is a counterpart of the Schwarzschild black hole in GR, has been obtained by Kehagia and Sfetsos \cite{KS}. This solution is usually known as the KS black hole. Then, in the slow rotation approximation, the black hole solution in the IR regime has been obtained in \cite{r1}--\cite{r2}. In the literature, many physical aspects of KS black hole have already been studied \cite{k1}--\cite{k10}. Moreover, for cosmological implications of Ho\v{r}ava-Lifshitz gravity see for instance Refs.\cite{c0}--\cite{c12}.

The shadows and rings of the KS black hole surrounded by a thin accretion disk have been studied in \cite{sh1}. However, the optical appearance of the KS black hole surrounded by spherical accretion flow has not yet been studied. So, in the present work, we consider the KS black hole surrounded with static/infalling accretion flows and discuss the effects of Ho\v{r}ava parameter and spherical accretion on the observed appearance of the black hole.

The paper is organized as follows. In section~\ref{2-KS BH}, after a brief review of KS black holes, we discuss the photon trajectories in the space-time of such black holes and investigate the effects of the Ho\v{r}ava parameter on them. Then we present the shadow images of KS black hole with static and infalling spherical accretion flows in section~\ref{3-rest} and section~\ref{4-infalling}, respectively. The paper ends with concluding remarks in section~\ref{5-conclusions}.

\section{KS black holes and trajectories of surrounding photons}
\label{2-KS BH}
\subsection*{A. KS geometry}
In the ADM formalism of Ho\v{r}ava-Lifshitz gravity the four-dimensional metric is parameterized as \cite{c0}
\begin{equation}
ds^2=-N^2c^2dt^2+g_{ij}(dx^i+N^idt)(dx^j+N^jdt),
\label{m1}
\end{equation}
where $N$, $N_{i}$ and $g_{ij}$ are the lapse function, the shift function and three-dimensional spatial metric, respectively.

The action of the IR-modified Ho\v{r}ava gravity is
\begin{eqnarray}
{\cal S}&=&\int dtdx^3\sqrt{g}N\left[\frac{2}{\kappa^2}(K_{ij}K^{ij}-\lambda K^2)-\frac{\kappa^2}{2\nu^4}C_{ij}C^{ij}+\frac{\kappa^2\mu}{2\nu^2}\epsilon^{ijk}R^{(3)}_{il}\nabla_{j}R^{(3)l}_{k}\right.\nonumber\\
&-&\left.\frac{\kappa^2\mu^2}{8}R^{(3)}_{ij}R^{(3)ij}+\frac{\kappa^2\mu^2}{8(3\lambda -1)}\left(\frac{4\lambda -1}{4}(R^{(3)})^2-\Lambda_{W}R^{(3)}+3\Lambda_{W}^2\right)+\frac{\kappa^2\mu^2\tilde{\omega}}{8(3\lambda -1)}R^{(3)}\right]
\label{1}
\end{eqnarray}
where $\mu$, $\nu$, $\lambda$, $\kappa$, $\tilde{\omega}$ and $\Lambda_{W}$ are constant parameters. $R^{(3)}$ is the three-dimensional curvature scalar for $g_{ij}$ and the extrinsic curvature, $K_{ij}$, is given by
\begin{equation}
K_{ij}=\frac{1}{2N}(\dot{g}_{ij}-\nabla_{i}N_{j}-\nabla_{j}N_{i}),
\label{2}
\end{equation}
where the dot represents a derivative with respect to $t$ and $\nabla_{i}$ denotes the covariant derivative with respect to the spatial metric $g_{ij}$. $C^{ij}$ is the Cotton tensor, reads as
\begin{equation}
C^{ij}=\epsilon^{ikl}\nabla_{k}\left(R^{(3)j}_{l}-\frac{1}{4}R^{(3)}\delta^{j}_{l}\right).
\label{3}
\end{equation}
Now, we consider the static and spherically symmetric metric as
\begin{equation}
ds^2= -g_{tt}(r)dt^2+g_{rr}(r)dr^2+r^2 d\theta^2+r^2 \sin^2\theta d\varphi^2,
\label{5}
\end{equation}
where $g_{tt}(r)$ and $g_{rr}(r)$ are functions of the radial coordinate $r$. In the specific case of $\lambda=1$, which reduces to the Einstein-Hilbert action in IR limit, the solution of the vacuum field equations can be obtained as follows
\begin{equation}
-g_{tt}(r)=1/g_{rr}(r)=f(r)=1+(\tilde{\omega}-\Lambda_{W}) r^2-\sqrt{r[\tilde{\omega}(\tilde{\omega}-2\Lambda_{W})r^3+\beta]},
\label{m2}
\end{equation}
where $\beta$ is an integration constant. By considering $\beta=4\tilde{\omega}M$ and $\Lambda_{W}=0$, the KS asymptotically flat solution is given by \cite{KS}
\begin{equation}
f(r)=1+\tilde{\omega} r^2\left[1-\left(1+\frac{4M}{\tilde{\omega} r^3}\right)^{1/2}\right],
\label{6}
\end{equation}
where $M$ is the mass of the black hole and $\tilde{\omega}$ is the Ho\v{r}ava-Lifshitz parameter. By rearranging the parameter $\tilde{\omega}$ as $\tilde{\omega}=\frac{1}{2\omega^2}$, one can rewrite the metric function in the following form \cite{sh1}
\begin{equation}
f(r)=1+\frac{r^2}{2\omega^2}\left(1-\sqrt{1+\frac{8M\omega^2}{r^3}}\right).
\label{07}
\end{equation}
The radius of the outer and inner horizons can be found by solving $f(r)=0$
\begin{equation}
r_{\pm}=M[1\pm\sqrt{1-\omega^2/M^2}].
\label{7}
\end{equation}
As is clear, for the existence of black hole solution a constraint $\omega/M\leq1$ should be imposed and for the extremal black hole, $r_+=r_-$, which corresponds to the case $\omega/M=1$. Also, in the limit of $\omega\rightarrow0$ the above metric becomes the static solution in GR which is described by the Schwarzschild metric. Note the parameter $\omega$ always takes the positive values. Thus, for the range $0<\omega<1$, the behavior of the inner horizon $r_-$ and the event horizon $r_+$,  is plotted in the left panel of Fig.~\ref{potential-radius}. The points at the beginning of each curve denote the corresponding values to the Schwarzschild solution.

\subsection*{B. Null trajectory around KS black hole}
\begin{figure}[H]
\centering
\includegraphics[width=3.0in]{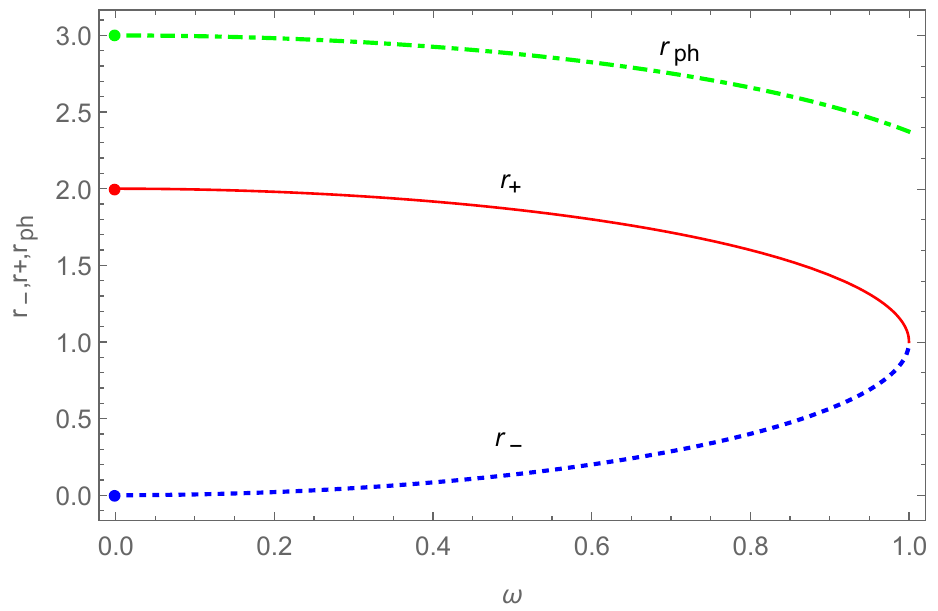}
\includegraphics[width=3.0in]{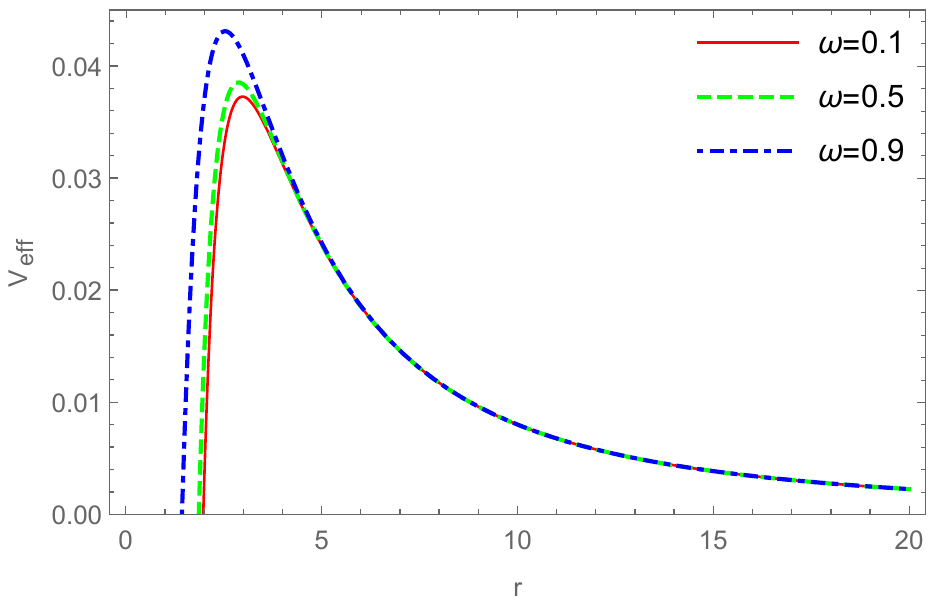}
\caption{\footnotesize Left panel: The photon radii $r_{\rm ph}$, $r_{-}$ and $r_{+}$ as a function of Ho\v{r}ava parameter for $M = 1$. Right panel: The effective potential $V_{\rm eff}$ as a function of radius for different values of Ho\v{r}ava parameter.}
\label{potential-radius}
\end{figure}

The trajectory of null geodesics in the space-time of KS black hole can be obtained using the Euler-Lagrange equations. Without loss of generality, we restrict ourselves to the equatorial plane, $\theta=\frac{\pi}{2}$. Since the metric coefficients of KS black hole are independent of $t$ and $\varphi$ coordinates, there are two constants of motion correspond to the energy $E$ and angular momentum $L$ of photons
\begin{equation}
E=f(r)\dot{t},
\label{mm8}
\end{equation}
\begin{equation}
L=r^2\dot{\varphi}.
\label{mm9}
\end{equation}
Now, taking the condition $2{\cal L}=g_{\mu\nu}\dot{x}^{\mu}\dot{x}^{\nu}=0$ for null geodesics and using above equations, we find the equations of photon motion around a KS black hole as
\begin{equation}
\dot{t}=\frac{E}{f(r)},
\label{8}
\end{equation}
\begin{equation}
\dot{\varphi}=\frac{L}{r^2},
\label{9}
\end{equation}
\begin{equation}
\dot{r}^2 = E^2 - V_{\rm eff}(r),
\label{10}
\end{equation}
where the effective potential is as follows
\begin{equation}
V_{\rm eff}(r)=\frac{L^2}{r^2}f(r)=\frac{L^2}{r^2}\left[1+\frac{r^2}{2\omega^2}\left(1-\sqrt{1+\frac{8M\omega^2}{r^3}}\right)\right].
\label{11}
\end{equation}
We have displayed the effective potential for different values of the Ho\v{r}ava parameter, $\omega$, in the right panel of Fig.~\ref{potential-radius}. As one can see, the peak of the potential increases with $\omega$. Note that for non-radial geodesics, it is convenient to set $L=1$ and thus we plot the figure for this value of the angular momentum.

Next, we focus on the photon motion in the vicinity of KS black hole. Combining equations (\ref{9}) and (\ref{10}) the differential equation governing the light rays trajectory can be obtained as
\begin{equation}
\left(\frac{dr}{d\varphi}\right)^2 = \frac{r^4}{b^2}-r^2f(r)
\label{12}
\end{equation}
where the impact parameter is defined as $b\equiv\frac{L}{E}$. In particular, for the photons with critical value of the impact parameter, $b = b_{\rm ph}$, an unstable
circular orbit occurs at the maxima of the effective potential at $r = r_{\rm ph}$, known as the photon sphere \cite{Claudel}. To study circular orbits with constant radius $r=r_{\rm ph}$ from equation (\ref{10}) we have
\begin{equation}
V_{\rm eff}(r_{\rm ph}) = E_{\rm ph}^2,\hspace {0.5 cm} V^{'}_{\rm eff}(r_{\rm ph}) = 0,
\label{14}
\end{equation}
where prime denotes differentiation with respect to the radial coordinate $r$. Use of equation (\ref{11}) leads to the following relation
\begin{equation}
rf'(r)-2f(r)=0
\label{15}
\end{equation}
which gives the radius of unstable photon circular orbits as
\begin{equation}
r_{\rm ph}=2\sqrt{3}M\cos\left[\frac{1}{3}\cos^{-1}\left(\frac{-4\omega^2}{3\sqrt{3}M^2}\right)\right].
\label{15}
\end{equation}
The dependence of the radius of the photon sphere $r_{\rm ph}$ on the parameter $\omega$ is also plotted in the right panel of Fig.~\ref{potential-radius}, showing that $r_{\rm [h}$ is a decreasing function of the Ho\v{r}ava parameter. Moreover, we see that in the limiting case $\omega\rightarrow0$, $r_{\rm ph}=3M$ which is the radius of the unstable circular photon orbit for the Schwarzschild black hole. The impact parameter of the photon sphere is also given by
\begin{equation}
b_{\rm ph} = \frac{r_{\rm ph}}{\sqrt{f(r_{\rm ph})}}=\left(\frac{1}{r_{\rm ph}^2}+\frac{1-\sqrt{1+\frac{8M\omega^2}{r_{\rm ph}^3}}}{2\omega^2}\right)^{-\frac{1}{2}},
\label{n17}
\end{equation}
which for an asymptotically flat space-time with a metric in the form (\ref{5}) is equal to the radius of the black hole shadow. The results of the radii of the event horizon $r_+$, photon sphere $r_{\rm ph}$ as well as the impact parameter of the photon sphere $b_{\rm ph}$ are also presented in Table.~\ref{T1}.

\begin{table}[H]
\centering
\caption{\footnotesize  The values of the event horizon radius $r_+$, photon radius, $r_{\rm ph}$ and impact parameter, $b_{\rm ph}$ for different values of $\omega$. The first column with $\omega\rightarrow0$ corresponds to the Schwarzschild black hole.}
\begin{tabular}{l l l l l l l l l l l}
\hline
$$&$\omega\rightarrow0$&$\omega=0.1$& $\omega=0.3$&$\omega=0.5$&$\omega=0.7$&$\omega=0.9$\\ [0.5ex]
\hline

$r_{\rm ph}/M$  &3        &2.9956     &2.9592   &2.8820   &2.7524  &2.5393  \\
$b_{\rm ph}/M$  &5.1962  &5.1923     &5.1609   &5.0951   &4.9868  &4.9117  \\
$r_+/M$         &2       &1.9949     &1.9539   &1.8660   &1.7141  &1.4359 \\
$r_-/M$         &0&       0.0050     &0.0461   &0.1339   &0.2859  &0.5641 \\

\hline
\end{tabular}
\label{T1}
\end{table}

\begin{figure}[H]
\centering
\includegraphics[width=3.0in]{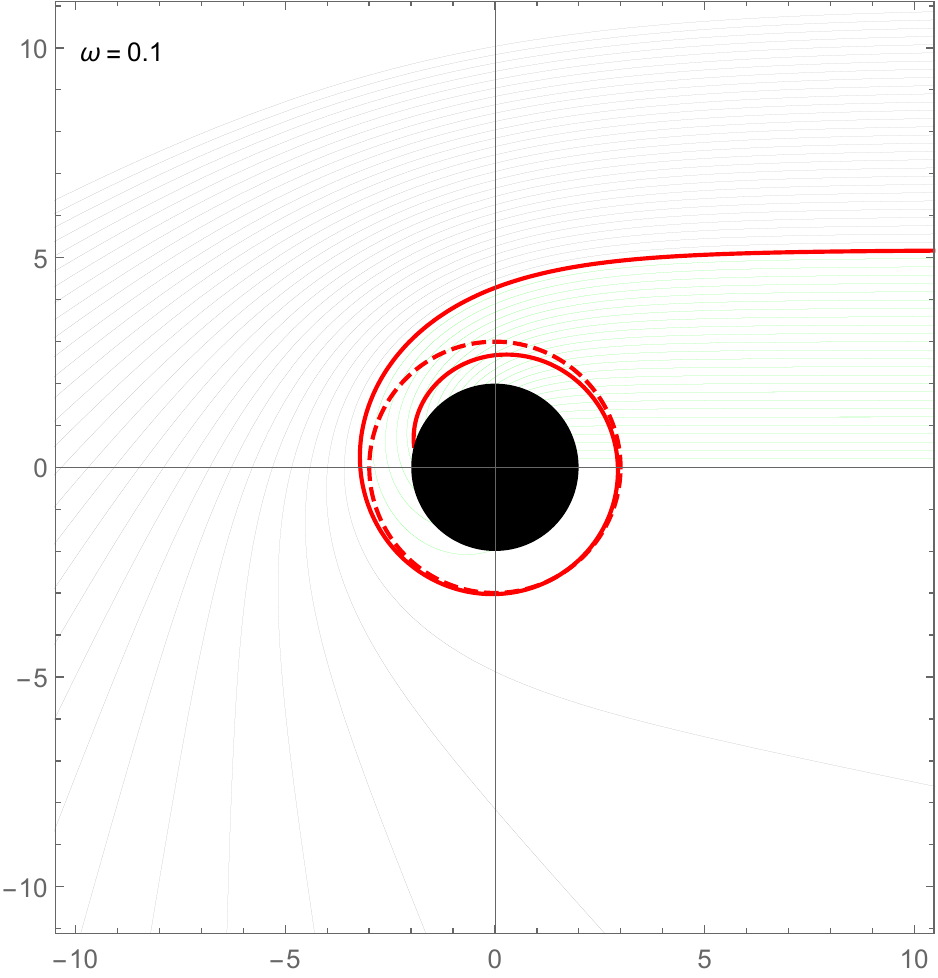}
\includegraphics[width=3.0in]{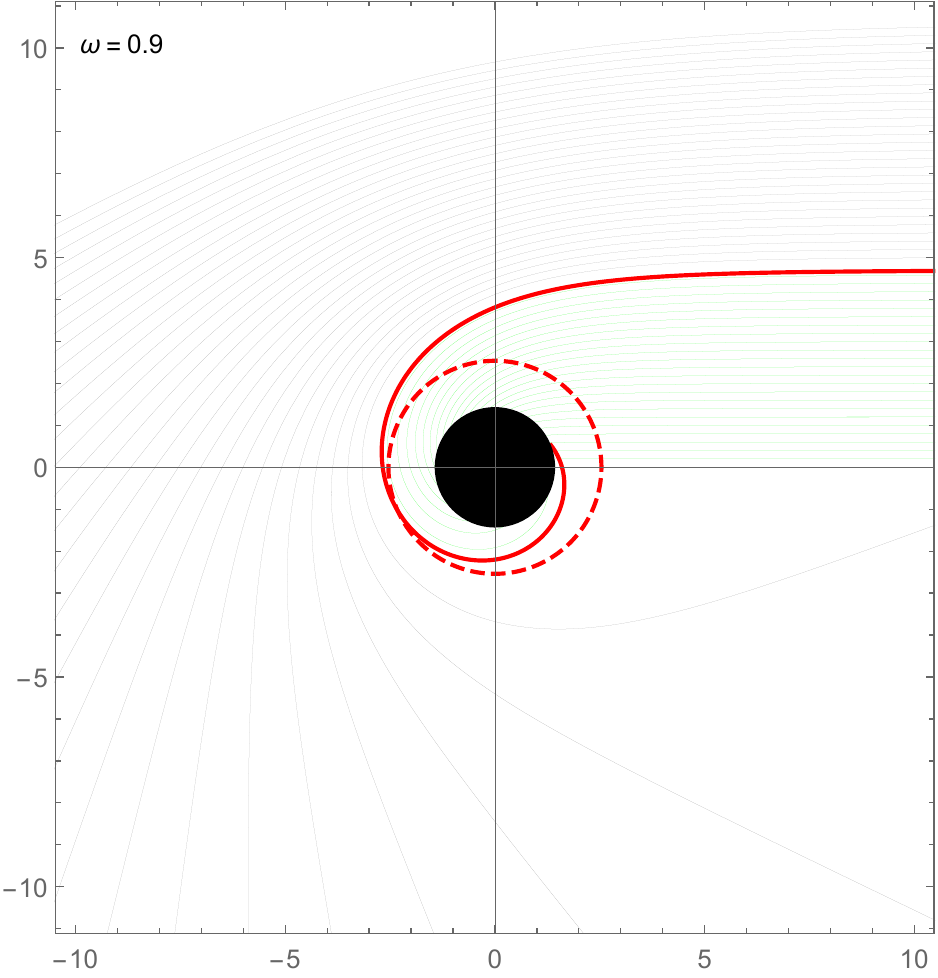}
\caption{\footnotesize The trajectory of light rays around a KS black hole with $\omega=0.1$ (left panel) and $\omega=0.9$ (right panel). The red, gray and green curves correspond to the trajectory of light rays with $b = b_{\rm ph}$, $b > b_{\rm ph}$ and $b < b_{\rm ph}$, respectively.}
\label{light-trajectory}
\end{figure}

From equation (\ref{12}), it is clear that the trajectory of light rays depends on the impact parameter $b$. The photons with small angular momentum, $b<b_{\rm ph}$, finally enter the black hole singularity, while photons with large angular momentum, $b>b_{\rm ph}$, will be deflected and pushed back to the distant observer. For critical values of impact parameter $b=b_{\rm ph}$, the photons swirl around the black hole on the photon sphere, with an infinite time. These photon trajectories are shown with green, gray and red curves in Fig.~\ref{light-trajectory}, respectively. The black disk also shows the event horizon surface. We have plotted the light trajectories for $\omega=0.1$ and $\omega=0.9$. As one can see, the event horizon radius and the radius of the photon sphere are smaller for a larger $\omega$. This is because the larger values of $\omega$ weakens the strength of gravity so that the instability area around KS black hole decreases and thus the photon radius takes smaller values.

\begin{figure}[H]
\centering
\includegraphics[width=3.2in]{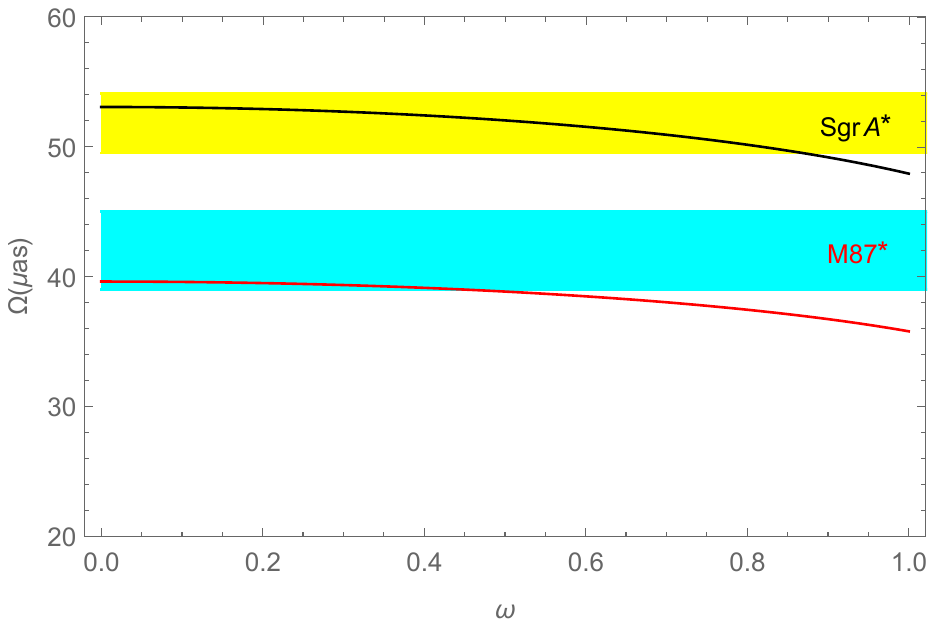}
\caption{\footnotesize Shadow diameter of the KS black hole as a function of Ho\v{r}ava parameter, $\omega$. The yellow and cyan regions are the experimental data of Sgr A* ($51.8 \pm 2.3$ $\rm \mu as$) and M87* ($42 \pm 3$ $\rm \mu as$) as reported by the EHT, respectively.}
\label{cn}
\end{figure}

Now, we aim to investigate the constraints on the Ho\v{r}ava parameter using the EHT observations of the shadow of M87* and Sgr A*. As one can see from equation (\ref{n17}), the shadow size is dependent on the Ho\v{r}ava parameter and thus from EHT observational data we can impose bounds on it. For a distant observer, the angular diameter $\Omega$ of the black hole shadow is given by \cite{co}
\begin{equation}
\Omega=\frac{2 b_{\rm ph}}{D},
\label{n1}
\end{equation}
where $D$ is the distance between the black hole and distant observer, and $b_{\rm ph}$ is obtained from equation (\ref{n17}). The above equation can be rewritten as
\begin{equation}
\left(\frac{\Omega}{\rm \mu as}\right) = \left(\frac{6.191165\times10^{-8}}{\pi}\frac{\gamma}{D/\rm Mpc}\right)\left(\frac{b_{\rm ph}}{M}\right),
\label{n2}
\end{equation}
with $\gamma$ is the mass ratio of the black hole to the Sun. From the reports released by EHT for the shadow of M87* the distance and mass correspond to $D = 16.8$ Mpc and $\gamma = 6.5\times10^{9}$, while for the shadow of Sgr A* these values are given by $D = 8.27$ kpc and $\gamma = 4.3\times10^{6}$, respectively \cite{A1}--\cite{A7}. In Fig.~\ref{cn}, we have plotted the diameter of the shadow of KS black hole as a function of Ho\v{r}ava parameter, using the data of M87* (red curve) and of Sgr A* (black curve). It is found that the $\Omega$ decreases with increase of $\omega$. The cyan and yellow regions are also the shadow diameters of M87* and Sgr A* reported by the EHT observations. As we observe, the KS black hole is able to describe the shadow size of Sgr A*, provided that the Ho\v{r}ava parameter is constrained to the values in range $0<\omega<0.8695$. We also constrain the parameter $\omega$ of the KS black hole as $0< \omega <0.4504$ for M87*.

\section{KS black hole with gas at rest}
\label{3-rest}
Now, we are going to study the optical appearance of KS black hole surrounded by spherical accretion flow, which is assumed to be optically thin. We consider two cases of accretion: a static spherical accretion model and an infalling spherical accretion.

First, we discuss the shadow image and photon sphere in the background of KS black hole surrounded by static spherical accretion flow. By integrating specific emissivity along the photon path $\gamma$, one can obtain the observed intensity of photons for a distant observer as follows \cite{flow1}--\cite{flow2}
\begin{equation}
I(\nu_{\rm o}) = \int_{\gamma} g^3 j(\nu_{\rm e}) dl_p,
\label{16}
\end{equation}
where $g \equiv \frac{\nu_{\rm o}}{\nu_{\rm e}}$ is gravity redshift factor, $\nu_{\rm e}$ is the radiated photon frequency, $j(\nu_{\rm e})$ is the emissivity per unit volume in the rest frame of the emitter, and $dl_{\rm p}$ is the infinitesimal proper length. For spherically symmetric space-time (\ref{5}) the redshift factor is $g = f(r)^{1/2}$. By considering monochromatic emission with rest-frame frequency $\nu_{\rm s}$ and the emission radial profile as $1/r^2$ \cite{flow2}, the specific emissivity is given by
\begin{equation}
j (\nu_{\rm e}) \propto \frac{\delta(\nu_{\rm e}-\nu_{\rm s})}{r^2}.
\label{17}
\end{equation}
Also, the proper length measured in the rest frame of the emitter for the KS black hole is
\begin{equation}
dl_{\rm p} = \sqrt{\frac{1}{f(r)}dr^2+r^2 d\varphi^2} = \sqrt{\frac{1}{f(r)}+r^2 \left(\frac{d\varphi}{dr}\right)^2} dr.
\label{18}
\end{equation}
Then, using equations (\ref{16})-(\ref{18}) and equation (\ref{12}), the photon intensity observed by a distant observer can be expressed as
\begin{equation}
I (\nu_{\rm o}) = \int_{\gamma}\frac{f(r)}{r^2}\sqrt{1+\frac{b^2 f(r)}{r^2-b^2 f(r)}} dr.
\label{19}
\end{equation}

\begin{figure}[H]
\centering
\includegraphics[width=3.0in]{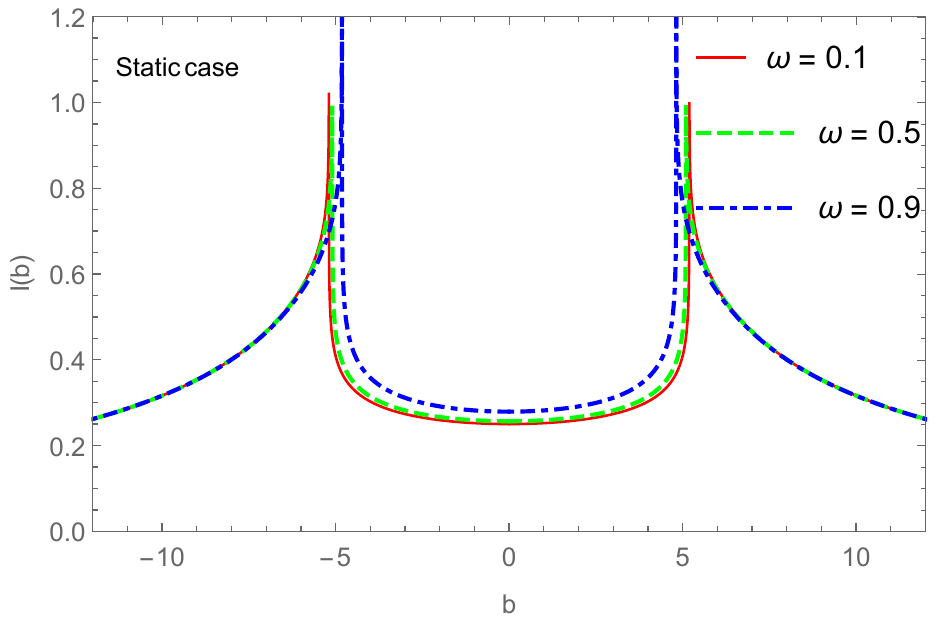}
\includegraphics[width=3.1in]{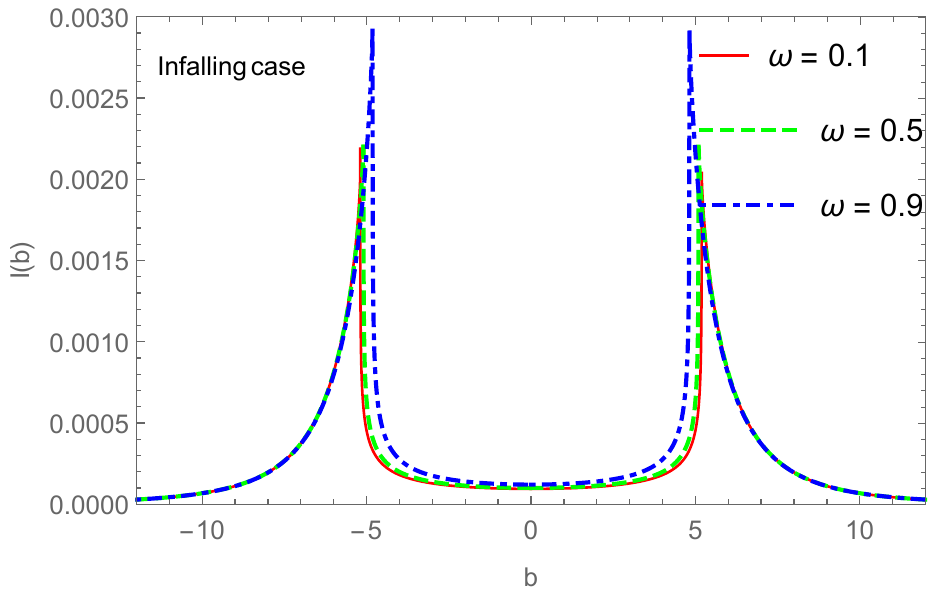}
\caption{\footnotesize The intensity profile as a function of impact parameter for KS black hole with gas at rest (left panel) and with radially infalling gas (right panel) for different values of $\omega$ and $M = 1$.}
\label{intensity}
\end{figure}

By substituting $f(r)$ from equation (\ref{07}), the specific intensity of KS black hole shadow and photon rings can be calculated. The observed specific intensity $I_{\rm o}$ is a function of both the impact parameter $b$ and the Ho\v{r}ava parameter $\omega$. We plot the behavior of intensity as a function of impact parameter for different values of $\omega$ in the left panel of Fig.~\ref{intensity}. As the figure shows, by increasing the Ho\v{r}ava parameter $\omega$ the light intensity of the shadow increases too. According to the figure the observed light intensity sharply increases with the increase of impact parameter and reaches its maximum value at $b_{\rm ph}$, due to the fact that the light rays round the black hole several times on the photon sphere, then the light intensity gradually decreases and takes its minimum value.

\begin{figure}[H]
\centering
\includegraphics[width=1.85in]{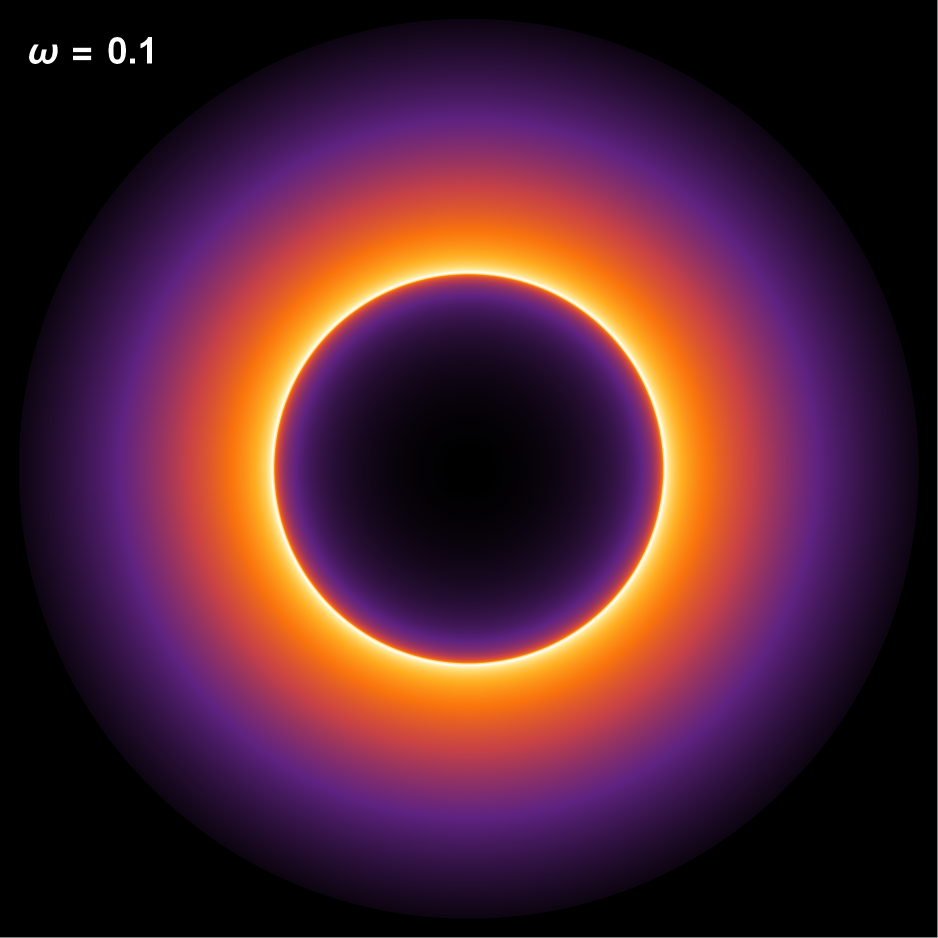}
\includegraphics[width=0.20in]{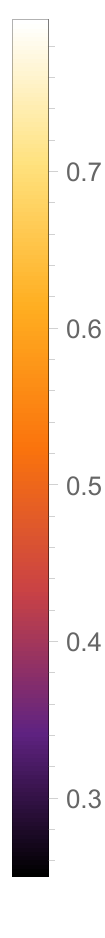}
\includegraphics[width=1.85in]{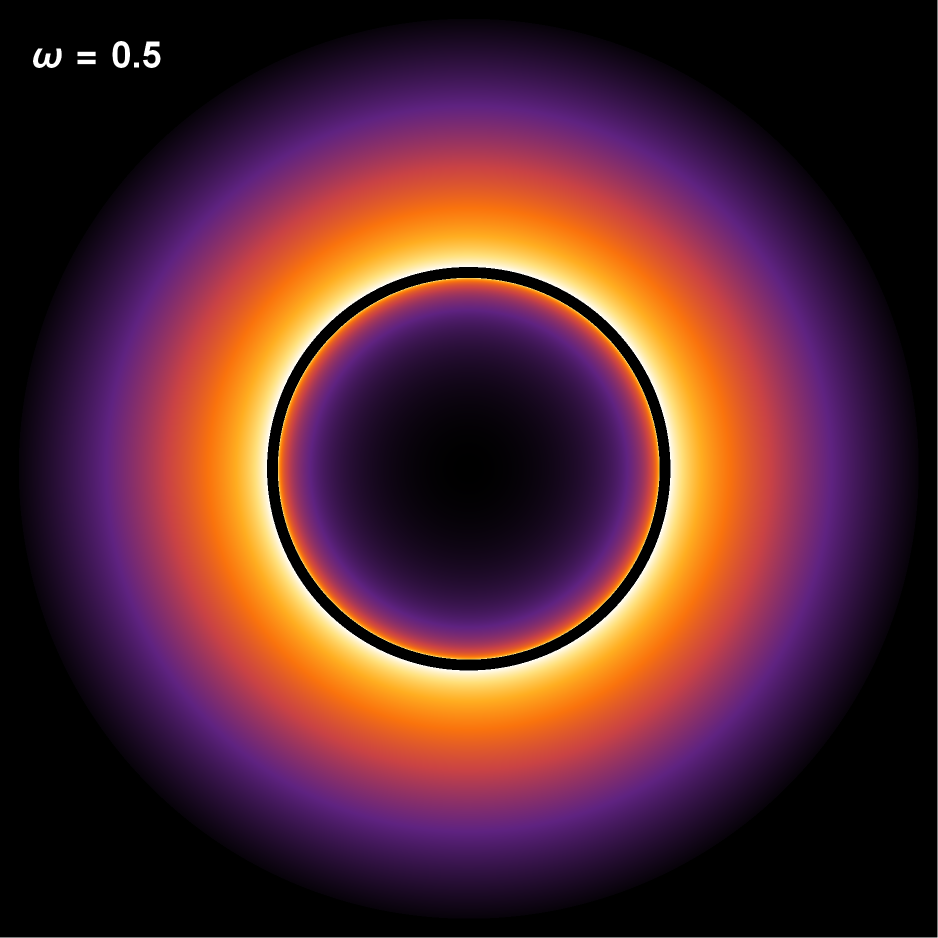}
\includegraphics[width=0.20in]{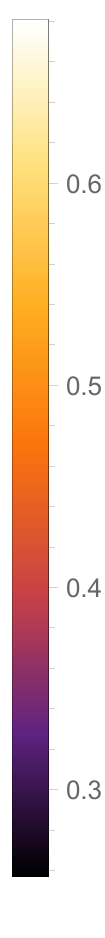}
\includegraphics[width=1.85in]{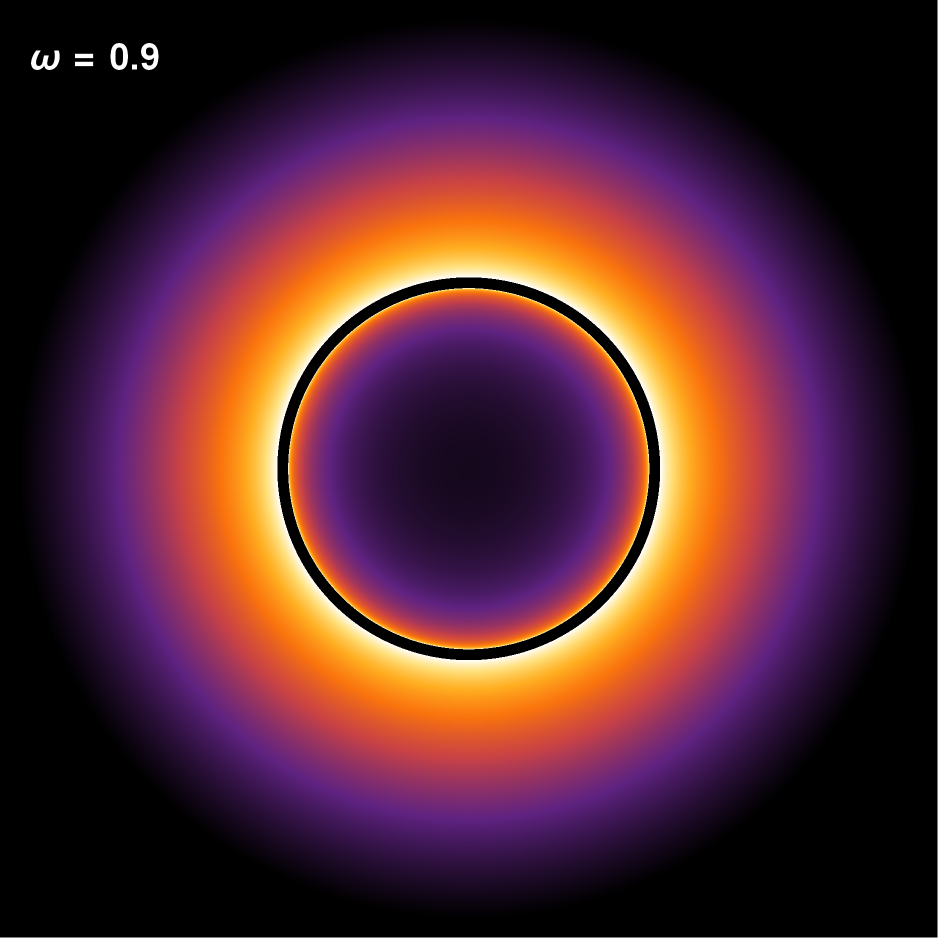}
\includegraphics[width=0.20in]{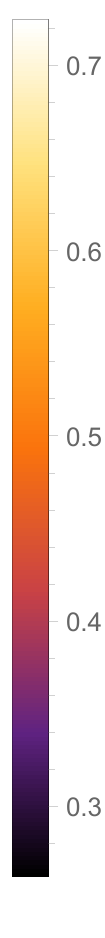}
\caption{\footnotesize The shadow and photon rings for KS black hole with gas at rest with different values of $\omega$ and $M = 1$.}
\label{shadow-static}
\end{figure}

Furthermore, the KS black hole shadow cast in the $(x,y)$ plane for the static spherical accretion is plotted in the Fig.~\ref{shadow-static}. In each panel, the central disk with a faint of luminosity is the black hole shadow and the brightest ring around the shadow shows the photon ring. As can be seen, with the increase of the parameter $\omega$, the radius of the shadow decreases with a smaller photon ring which is in agreement with Table.~\ref{T1}. Also, according to the figure, by increasing $\omega$ the specific intensity increase since, with the increase of $\omega$ the strength of gravitational field of KS black hole decreases which leads to lower light rays being trapped by KS black hole and thus in this region higher luminosity of shadow and photon rings being observed. We also conclude that only the KS black holes with larger values of $\omega$ have significant deviation from the Schwarzschild black hole with $\omega=0$ in GR.

\section{KS black hole with infalling gas}
\label{4-infalling}
In this section we consider the KS black hole surrounded by radial infalling spherical accretion flow. This model is more realistic than the static accretion model since most of accretions are moving in reality. In this scenario, the redshift factor is related to the velocity of accretion flow as
\begin{equation}
g = \frac{k_{\alpha}u_{\rm o}^{\alpha}}{k_{\beta}u_{\rm e}^{\beta}},
\label{20}
\end{equation}
where $k_{\mu}$ is the photon four-velocity, $u^{\mu}_{\rm o}=(1,0,0,0)$ is the distant observer four-velocity, and $u^{\mu}_{\rm e}$ is the four-velocity of the infalling spherical accretion flow. Based on equations (\ref{8})-(\ref{10}), we know that $k_t=1/b$ is a constant and $k_r$ can be obtained by the equation $k_{\mu}k^{\mu}=0$. Therefore,
\begin{equation}
\frac{k_r}{k_t} = \pm\sqrt{\frac{1}{f(r)}\left(\frac{1}{f(r)}-\frac{b^2}{r^2}\right)},
\label{21}
\end{equation}
where the sign + (-) corresponds to the case that the photon approaches (away from) the black hole. Also, the four-velocity of the infalling accretion is given by
\begin{eqnarray}
&u_{\rm e}^{t}& = \frac{1}{f(r)},\nonumber\\
&u_{\rm e}^{r}&= -\sqrt{1-f(r)},\nonumber\\
&u_{\rm e}^{\theta}&=u_{\rm e}^{\varphi}=0.
\label{22}
\end{eqnarray}
Thus, the redshift factor in equation (\ref{20}) can be obtained as
\begin{equation}
g = \frac{1}{u_{\rm e}^t+\left(\frac{k_r}{k_t}\right)u_{\rm e}^r}.
\label{23}
\end{equation}
In addition, the proper length is
\begin{equation}
dl_{\rm p} = k_{\alpha}u^{\alpha}_{\rm e}ds = \frac{k_t}{g \mid k_r\mid}dr,
\label{24}
\end{equation}
where $s$ is the affine parameter along the photon path. Assuming that the specific emissivity has the same form as equation (\ref{17}), the specific intensity $I(\nu_{\rm o})$ in the case of infalling spherical accretion can be expressed as
\begin{equation}
I(\nu_{\rm o}) \propto \int \frac{g^3k_t dr}{r^2 |k_r|}.
\label{25}
\end{equation}

\begin{figure}[H]
\centering
\includegraphics[width=1.85in]{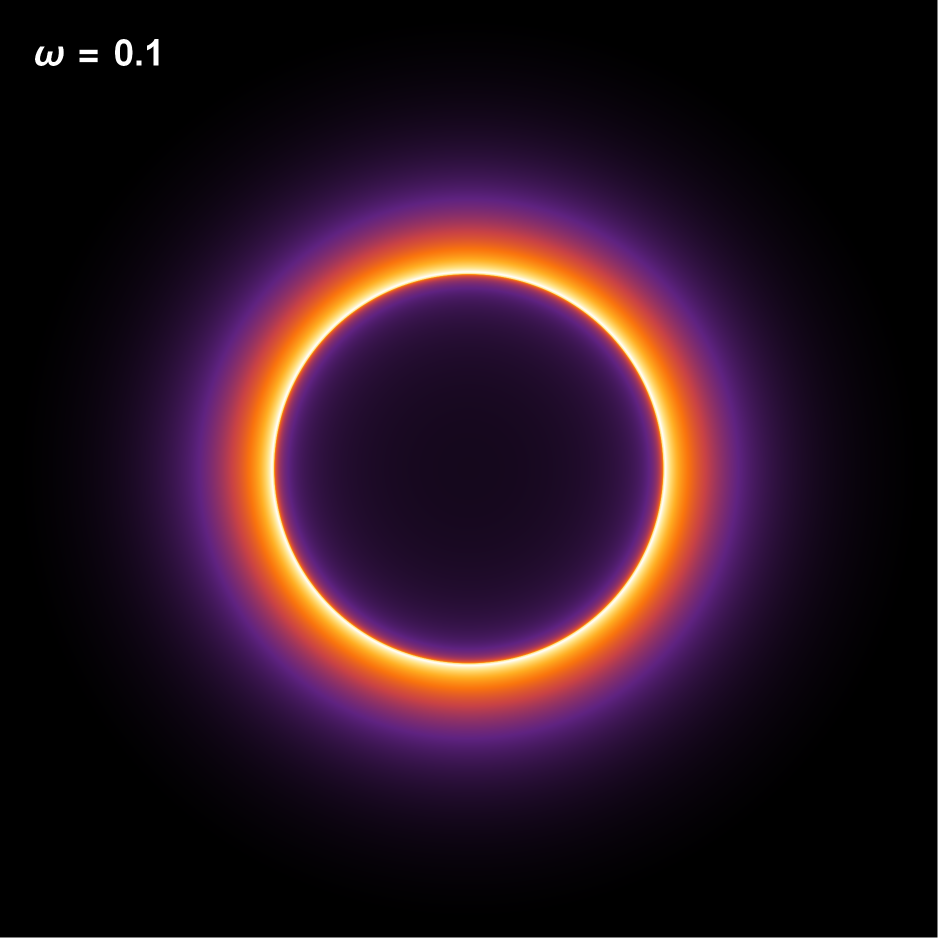}
\includegraphics[width=0.30in]{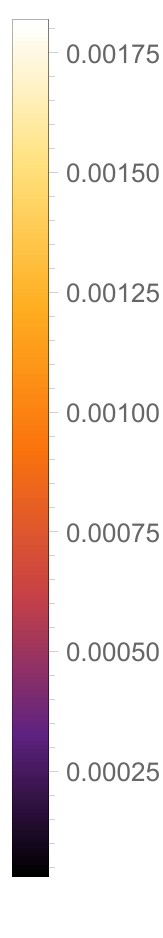}
\includegraphics[width=1.85in]{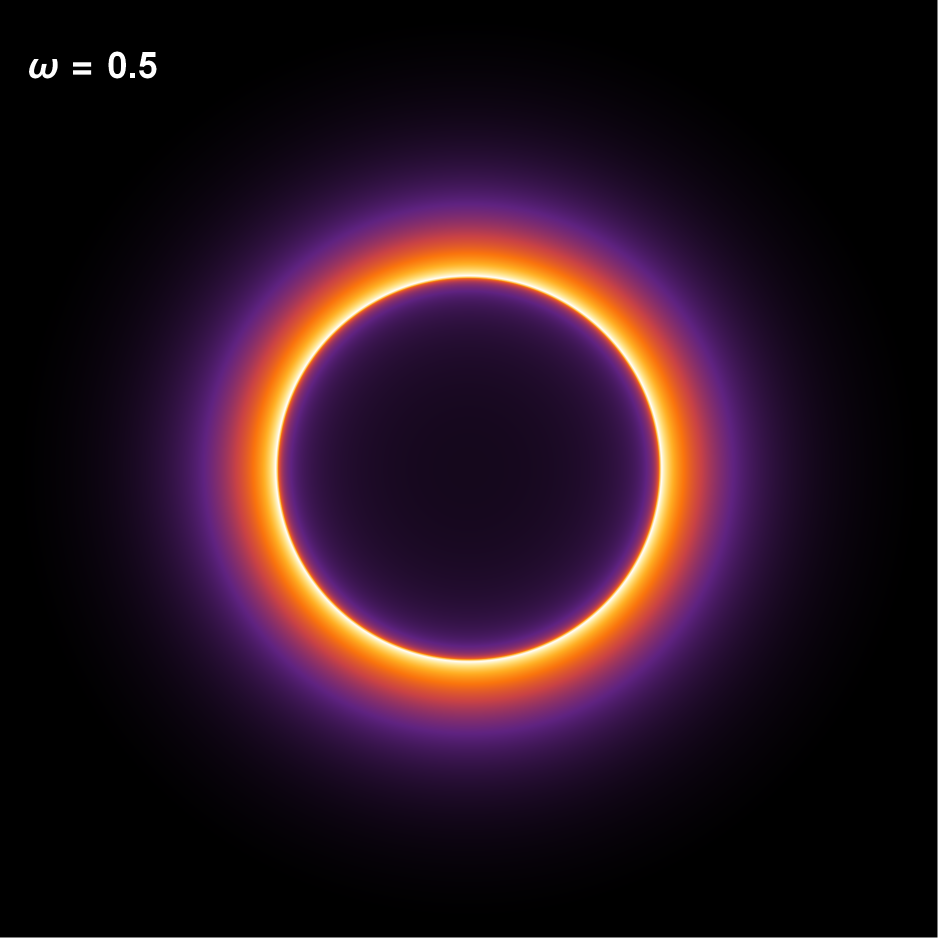}
\includegraphics[width=0.28in]{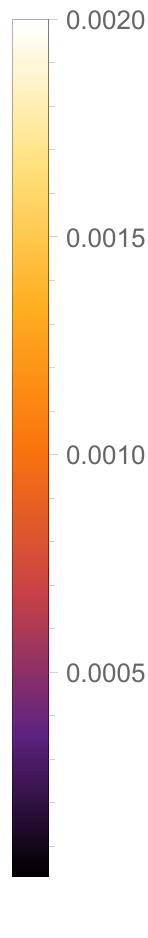}
\includegraphics[width=1.85in]{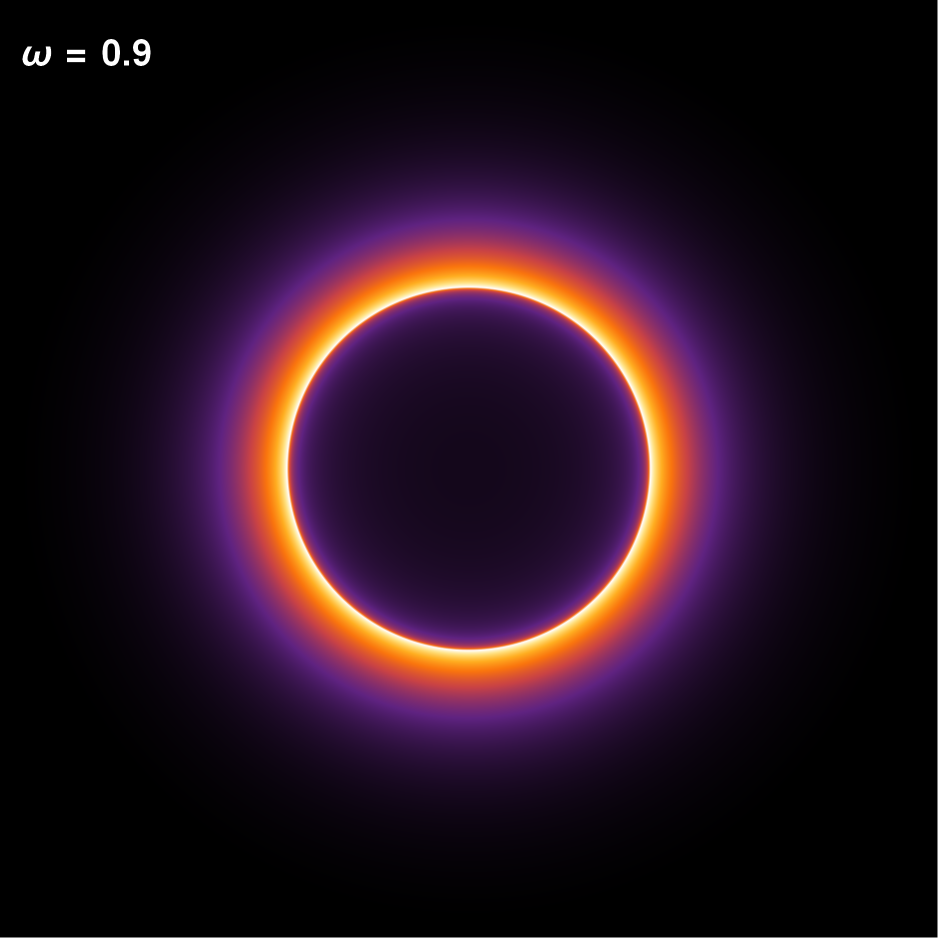}
\includegraphics[width=0.28in]{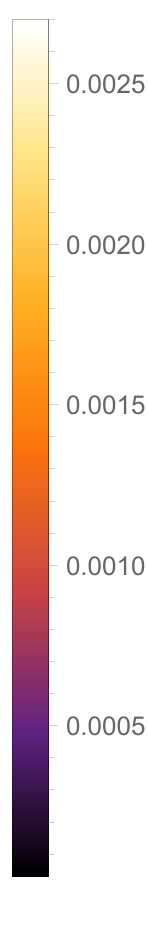}
\caption{\footnotesize The shadow and photon rings for KS black hole with radially infalling gas with different values of $\omega$ and $M = 1$.}
\label{shadow-fall}
\end{figure}

Similar to the static model, the observed intensity for the infalling spherical accretion as a function of impact parameter is plotted in the right panel of Fig.~\ref{intensity}. As can be seen, the maximum value of intensity is at $b_{\rm ph}$ but the peak of intensity for infalling case is smaller than the static one. The numerical results of intensity shows that the observed intensity increases with increasing $\omega$. We also present the two-dimensional plot of shadow cast in Fig.~\ref{shadow-fall}, showing that the radius of shadow and photon ring decreases with increase of parameter $\omega$, while the shadow and photon rings luminosities increase. Comparing Fig.~\ref{shadow-static} for the static case with Fig.~\ref{shadow-fall} for the infalling model shows that the shadow region for the infalling accretion flow is darker than that of the static case which is caused by the Doppler effect of the infalling matter. Note that the observed shadow size depend only on the space-time geometry, while the luminosities of both the shadow and photon rings are also affected by the accretion flow property.

\section{Conclusions}
\label{5-conclusions}
In this paper, we studied the optical appearance of the KS black holes surrounded by spherical accretion flow. First, we obtained the size of the event horizon radius, the black hole shadow and photon sphere for different values of Ho\v{r}ava parameter $\omega$, and found that the larger the $\omega$ parameter is, the smaller the radius of the event horizon,  photon sphere and shadow will be. We also obtain the constraint on the parameter $\omega$ using the observed shadow diameters measured by the EHT, which are $0<\omega<0.8695$ for Sgr A*, and $0<\omega<0.4504$ for M87*. Then, we study the observed intensity and the luminosity of KS black hole shadows and photon rings by assuming that the black hole is surrounded by static/infalling spherical accretion flow. For both scenarios, we found that the luminosities of the shadow and photon rings of KS black holes increase with the increase of Ho\v{r}ava parameter. This is because by increasing $\omega$, the impact parameter decreases which means that the photons gain more kinetic energy and thus not easily captured by KS black hole which leads to increase the luminosity. Therefore, it can be argued that this observational appearance may offer a way to distinguish a KS black hole in Ho\v{r}ava-Lifshitz gravity from a Schwarzschild black hole in GR.

\section*{Acknowledgments}
The work of Mohaddese Heydari-Fard is supported by the Iran National Science Foundation (INSF) and the Research Council of Shahid Beheshti University under research project No. 4016024.

\end{document}